\documentclass[iop]{emulateapj}
\usepackage{natbib}
\usepackage{apjfonts}

\slugcomment{To appear in the Astrophysical Journal Letters}

\newcommand{\rv}{{\rm v}}
\newcommand{\vv}{\vec{v}}
\newcommand{\dd}{\mathrm{d}}
\newcommand{\tcross}{t_\mathrm{cross}}
\newcommand{\tcoll}{t_\mathrm{cont}}
\newcommand{\tdyn}{t_\mathrm{dyn}}

\shorttitle{Adiabatic Heating in Contracting Turbulent Fluids}
\shortauthors{Robertson \& Goldreich}

\begin{document}

\title{Adiabatic Heating of Contracting Turbulent Fluids}
    
\author{Brant Robertson\altaffilmark{1} and Peter Goldreich\altaffilmark{2}}

\altaffiltext{1}{Steward Observatory, University of Arizona, 933 North 
Cherry Avenue, Tucson, AZ 85721}
\altaffiltext{2}{California Institute of Technology, 
1200 East California Boulevard, Pasadena, CA 91125}

\begin{abstract}
Turbulence influences the behavior of many astrophysical systems, frequently 
by providing non-thermal pressure support through random bulk motions. 
Although turbulence is commonly studied in systems with
constant volume and mean density, 
turbulent astrophysical gases often expand or contract
under the influence of pressure or gravity. 
Here, we examine the behavior of
turbulence in contracting volumes using idealized models of compressed gases.
Employing numerical simulations and an analytical model, we identify a simple mechanism by which the
turbulent motions of contracting gases ``adiabatically heat'',
experiencing an increase in their random bulk velocities 
until the largest eddies in the 
gas circulate over a ``Hubble'' time of the contraction. Adiabatic heating 
provides a mechanism for 
sustaining turbulence in gases where no large-scale driving exists.
We describe this mechanism in detail and discuss some potential applications to
turbulence in astrophysical settings.
\end{abstract}


\section{Introduction}
\label{section:introduction}

Turbulence -- the bulk random motion of a gas or fluid -- 
is ubiquitous in astrophysics.
Turbulence can be generated 
by 
instabilities, including gravitational \citep{jeans1902a},
shear \citep{helmholtz1868a, kelvin1871a}, convective \citep{rayleigh1883a, taylor1950a}, 
and magnetorotational \citep{balbus1991a}.
For overviews, see 
\citet{elmegreen2004a} and \citet{mckee2007a}. 
Given its wide-spread importance,
turbulence remains a critical area for astrophysical research.

Supersonic turbulence has been studied
in great detail with numerical simulations.  For instance, 
supersonic isothermal turbulence exhibits a
lognormal density distribution, with a width that increases with 
the Mach number
\citep[e.g.,][]{vazquez-semadeni1994a,padoan1997a,
kritsuk2007a, lemaster2008a, federrath2010a, price2011a}.
The properties of supersonic isothermal turbulence 
appear to be independent of the simulation methodology 
\citep[e.g.,][]{kitsionas2009a,price2010a,bauer2011a}. Most
studies of turbulence have involved gases simulated
in a static volume, whereas astrophysical gases 
often expand or contract under the influence of 
pressure or gravity.  Little is currently
known about the detailed structure of expanding or contracting 
turbulent gases.

In this {\it Letter}, we examine the behavior of turbulence during the
contraction of a gas arising from pressure or self-gravity.
In Section \ref{section:simulations}, we use simulations to 
model 
contracting turbulence and demonstrate
that turbulence adiabatically heats during contraction provided the eddy
turnover time\footnote{While the term {\it eddy} accurately describes the 
vortices of incompressible turbulence, it is less
accurate for motions in compressible turbulence.  Lacking a better term, 
we nonetheless refer to the large scale motions in compressible turbulence 
as eddies. Similarly, the term {\it turnover time} is used to describe
the timescale of these motions.} is shorter than the contraction time. We term this 
mechanism ``adiabatic heating'', and in Section
\ref{section:model} we present an analytical model that successfully 
describes its behavior.
We discuss some 
potential astrophysical applications of adiabatic heating 
in Section \ref{section:discussion}, and summarize and conclude in 
Section \ref{section:summary}.

\begin{figure*}
\figurenum{1}
\epsscale{1}
\plotone{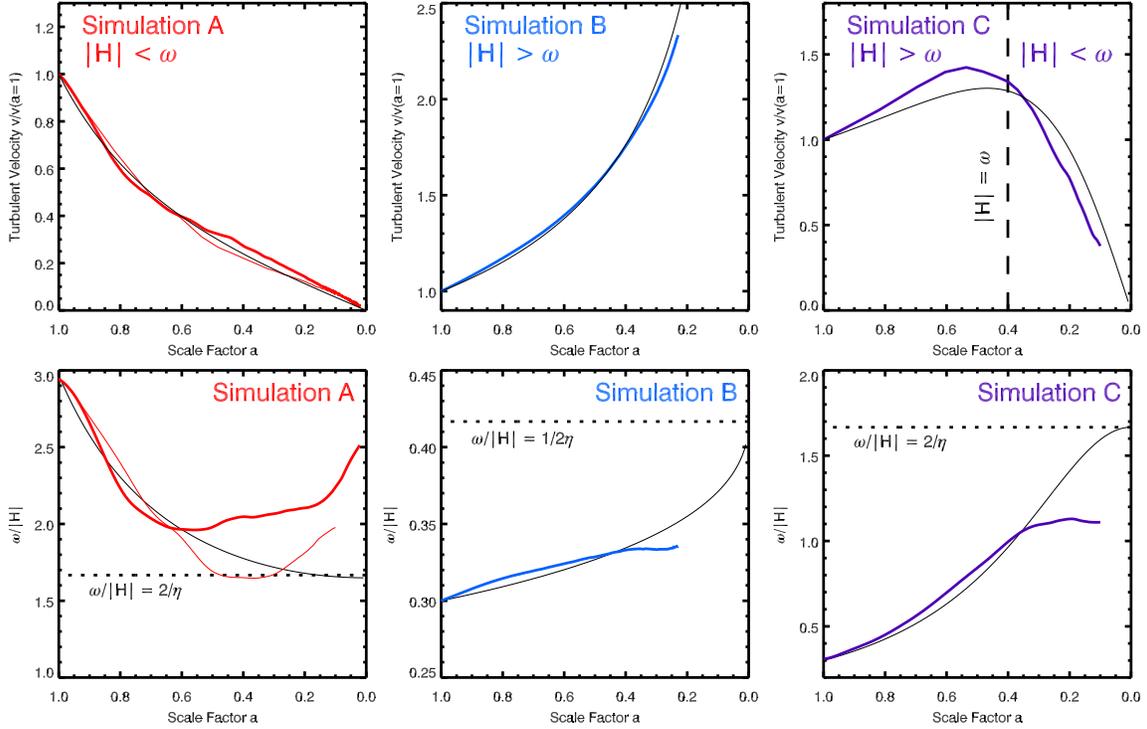}
\caption{\label{fig:mach_number}
Contraction of a turbulent gas and the adiabatic heating mechanism. 
Shown are the RMS turbulent velocities for contracting isothermal simulations with differing Hubble parameters (colored lines).
In an initially ``slow'' contraction (top left panel) where the contraction frequency 
$|H|$ is lower than
the eddy turnover frequency $\omega$,  large-scale eddies circulate and the 
turbulent cascade transfers energy to small scales where it dissipates.  Instead, if the 
contraction frequency is initially higher than the eddy turnover frequency (top center and right panels), the 
turbulence heats roughly adiabatically at first. As the ratio of eddy turnover and contraction frequencies in the simulations become
comparable, whether the turbulence heats or decays as the scale factor $a\to0$ depends on the 
evolution of $H(a)$.  In a dynamical contraction (top center panel) where $H\propto a^{-3/2}$ the turbulence continues to heat after $\omega$ tracks $|H|$,  whereas for constant $H$ (top right panel) it decays. The solid black lines in each panel show the evolution of turbulent velocities predicted by our
analytic adiabatic heating model (see Section \ref{section:model}).
The dotted lines indicate the model predictions for the synchronized ratio of $\omega/|H|$.  To indicate the 
effect of the forced turbulence initial
conditions
on the contracting turbulence, the Simulation A panels show two realizations (thick and thin lines).
}
\end{figure*}

\section{Numerical Simulations of Contracting Turbulence}
\label{section:simulations}

We use hydrodynamic simulations 
to study turbulence
in a contracting
background.
We model the contraction by
parameterizing the changing physical size $l(t)$  
and coordinate scale factor $a = l(t)/L$ of a cubic
volume of initial length $L$ 
through a ``Hubble''
parameter $H\equiv \dot{a}/a$ that may depend on time $t$. 
In terms of the proper coordinates within an
isotropically contracting volume, 
the Euler equations connecting 
derivatives of the 
density $\rho$, momentum $\rho\vv$, and pressure $p$
are altered by terms that depend on the 
Hubble parameter as
\begin{equation}
\label{eqn:density_derivative}
\frac{\partial \rho}{\partial t} = - \nabla\cdot(\rho\vv) -3H\rho 
\end{equation}
\begin{equation}
\label{eqn:momentum_derivative}
\frac{\partial \rho \vv}{\partial t} = -\nabla\cdot(\rho\vv\vv) - \nabla p -4H\rho\vv  
\end{equation}
\noindent
\citep[see, e.g., Section 9 of][]{peebles1980a}.
For an initially constant density and velocity gas
without dissipation, these terms give 
rise to two important scalings well known from cosmology: $\rho\propto a^{-3}$ 
and $v\propto a^{-1}$. 
Mass conservation dictates the density scaling, but 
the presence of dissipation (either through physical viscosity 
or numerically through the discretized form of Equation
\ref{eqn:momentum_derivative}) implies that the adiabatic velocity scaling does
not strictly hold
in the contraction of a 
turbulent gas. 
How the turbulent velocity evolves depends 
on how dissipation operates during the contraction, and simulations 
are required to provide a detailed description.

The simulations were performed
using a version of the 
magnetohydrodynamics code {\it Athena} \citep{stone2008a} modified to model 
contracting and expanding turbulent gases (see Equations \ref{eqn:density_derivative}
and \ref{eqn:momentum_derivative}, and below).  
{\it Athena} is a 
grid code 
based on the \citet{godunov1959a} method.
The calculations use 
piecewise parabolic reconstruction \citep{colella1984a} to extrapolate initial 
states for the Riemann problem between cells and compute final states 
using an exact solver \citep[][]{toro1999a}. Cell-averaged conserved 
quantities are updated using unsplit methods \citep[][]{gardiner2008a}. 
Our modifications to {\it Athena} include a Runge-Kutta integrator to evolve
a differential equation for the scale factor $a$ that depends on the
possibly time-dependent Hubble parameter $H$.  

The initial conditions are snapshots of driven 
isothermal turbulence (with sound speed $c_s=1$ and mean density $\bar{\rho}=1$) 
simulated on an $N=512^{3}$ resolution grid. 
The random forcing field 
is generated following
\citet{bertschinger2001a}, with power input into the two largest modes in 
the unit ($L=1$) periodic box \citep[e.g.,][]{kritsuk2007a}.  Driving at
intermediate scales produces similar results.
A Helmholtz decomposition in Fourier 
space removes the dilatational component and each forcing field is 
normalized to maintain an average Mach number $M\approx6$ 
when applied as an acceleration 
ten times per crossing time $\tcross \approx L/2Mc_{s}$. 
The driving is applied 
for ten crossing times and then terminated before the contraction initiates.
Although we simulate isothermal gases, results relevant for 
the adiabatic heating
mechanism originate from Equations \ref{eqn:density_derivative} and 
\ref{eqn:momentum_derivative} 
and should generalize to other adiabatic indices.

\begin{figure*}
\figurenum{2}
\epsscale{1}
\plotone{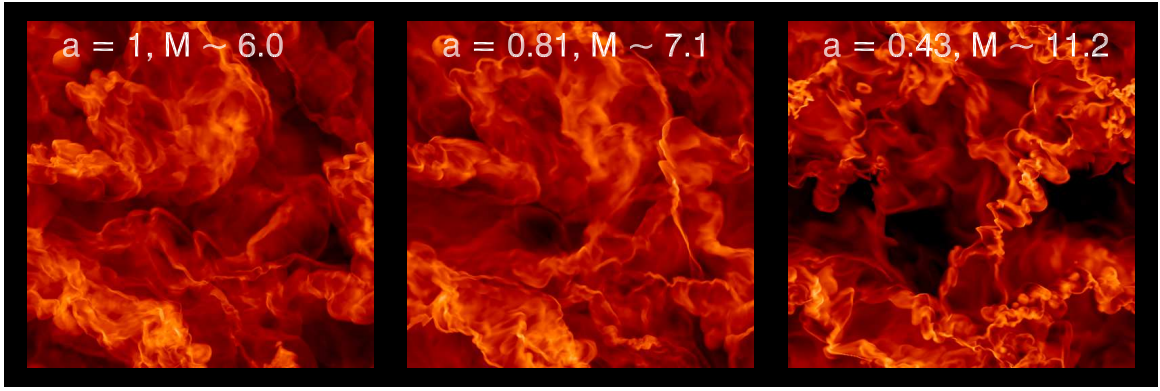}
\caption{\label{fig:dynamic_H}
Adiabatic heating of isothermal supersonic turbulence during contraction.
Shown is 
the logarithmic density distribution through a thin (8/512) slice of Simulation B
at three values of the scale factor $a$. 
Here, the Hubble parameter scales with the 
inverse dynamical time of the gas.
The density maximum in the color map of each panel is scaled by $a^{-3}$, and the dynamic 
range of each image is $10^4$. 
The turbulence adiabatically heats from the initial Mach number of $M\sim6$ to 
$M\sim11$ as the scale factor decreases by a factor of $\sim2$. The presence of 
limited dissipation breaks the perfect adiabatic scaling. The bulk properties of the 
gas in each panel, including the density distribution and intermittency, behave similarly to 
static-frame isothermal turbulent gases at the same Mach number. 
}
\end{figure*} 

\subsection{Models of Contraction}
\label{section:contraction}

Two rates characterize the isotropic contraction of a turbulent gas, the contraction frequency 
$|H| \equiv |\dot{a}/a|$, also called the Hubble parameter ($H<0$ for a contraction), 
and the eddy turnover frequency 
\begin{equation}
\label{eqn:eddy_frequency}
\omega\sim\frac{\rv}{aL},
\end{equation}
\noindent
where $\rv$ is the root-mean-squared (RMS) turbulent velocity (for isothermal turbulence $\rv\equiv M c_{s}$, where $M$ is the
typical Mach number and $c_{s}$ is the sound speed).
In our simulations, the values of $\rv(t=0)\approx6c_{s}$, 
$c_{s}=1$, and $L=1$ imply an initial
eddy turnover frequency $\omega(t=0) \approx 6$.  
To demonstrate the generality of the adiabatic heating mechanism we simulate 
three different scenarios for the time-dependent relation between $H$ and $\omega$:
\\\\
\noindent
{\bf Simulation A}: an initially ``slow exponential contraction'', with $H = $~constant.  
In this case the scale factor evolves 
as $a(t) = a_{0} \exp[H(t-t_{0})]$, where 
$a_{0}=1$ is the initial scale factor and $t_{0}$ is the time when the
contraction ensues.  We choose a constant Hubble parameter $H=-2$, such that the contraction
is slow (i.e., $|H|\ll\omega$) initially.  To characterize run-to-run variations, we perform
two such simulations differing only in their forced turbulence initial conditions.
\\\\
\noindent
{\bf Simulation B}: an initially ``fast dynamical contraction'', with $H \propto \sqrt{\bar{\rho}} $. 
In this case the contraction time $\tcoll\sim |H|^{-1}$
scales with the 
dynamical time $\tdyn\propto1/\sqrt{\bar{\rho}}$ set by the mean density $\bar{\rho}\propto a^{-3}$.
Starting with an
initial value $H(t=t_{0}) = H_{0}$, the Hubble parameter
varies with the scale factor as $H = H_{0} (a/a_{0})^{-3/2}$.
The scale factor decreases with time as $a(t) = a_{0}[3H_{0}(t-t_{0})/2 +1]^{2/3}$ (recall
that $H_{0}<0$).  We set $H_{0}=-20$ to induce an initially fast ($|H|\gg\omega$) contraction.
\\\\ 
\noindent
{\bf Simulation C}: an initially ``fast exponential contraction'', 
with $H = $~constant.  
In this case the scale factor evolves with the same time-dependence as in Simulation A,
but at a constant contraction frequency 
($|H|=20$) such that $|H|\gg\omega$ initially. 

\subsection{Simulation Results}
\label{section:results}

The properties of contracting turbulence evolve with decreasing scale factor $a$
in a manner that depends upon the ratio 
of the eddy turnover frequency to the contraction frequency (Figure \ref{fig:mach_number}).
The simulations demonstrate that if the contraction is slow ($|H|\ll\omega$ 
initially in Simulation A, left panels), the turbulent velocities decay, 
whereas if it is fast ($|H|\gg\omega$ initially in Simulations B and C, center and right panels), 
the turbulent velocities amplify.  In a slow contraction, large vortices circulate and nonlinear 
interactions transfer energy to smaller scales where it is dissipated before the box shrinks 
appreciably.  When the contraction is fast, energy bearing eddies are adiabatically 
compressed, dissipation primarily operates on small scales, and turbulent velocities increase.  
In each example in Figure \ref{fig:mach_number}, $\omega/|H|$ become comparable (bottom panels).  
From this trend we surmise that the
eddy turnover frequency may eventually 
``synchronize'' with the contraction frequency.  Our simulations provide a hint 
of this behavior, but the 
synchronized state is not well-explored.  Physical considerations
suggest the synchronization is stable since the eddies are compressed
on their circulation timescale, and the velocities of large eddies 
should hover around $\rv\sim |H|aL$. If true, as $a\to 0$, 
the velocities would decrease for constant $|H|$ as observed in 
Simulations A and C but continue to increase for $|H|\propto a^{-3/2}$ as seen in Simulation B.  

Figure \ref{fig:dynamic_H} shows 
the projected density 
distribution through a slice of Simulation B at three scale factors
during the contraction.  Initially, at $a=1$ (left panel) 
the gas has a turbulent velocity $\rv\equiv Mc_{s}\approx6c_{s}$,
a lognormal density distribution, and a velocity power spectrum characteristic of 
supersonic turbulence.  Compression during contraction heats the turbulence in the isothermal gas 
to Mach number
$M=7.1$ ($M=11.2$) by scale factor $a=0.81$ ($a=0.43$).  As the Mach number increases, the
width of the lognormal density distribution increases, the intermittency amplifies, 
and the velocity power spectrum steepens
much in the same way as driven, constant volume isothermal turbulence simulations behave as a function of 
Mach number \citep[e.g.,][]{price2011a}.  The increase in turbulent velocities arises from
the approximate inverse dependence of velocity on the scale factor during contraction. 
Since a turbulent cascade transfers large-scale power to small scales where it dissipates, the 
degree to which the turbulent velocity tracks $a^{-1}$ during the simulation
depends on the rate of energy transfer to small scales.

\section{Analytical Model of the Adiabatic Heating Mechanism}
\label{section:model}

We model the behavior of turbulent velocities during contraction by 
calculating the approximate time rate of change of the kinetic energy per unit mass in the 
gas, including two important terms. A term 
capturing the effects of adiabatic heating follows from noting that the 
adiabatic velocity scaling implies $\rv a=\rv_0$, where $\rv_0$ is a constant. Thus
\begin{equation}
\label{eqn:KE_AH}
\frac{\dd}{\dd t} \left(\frac{\rv^{2}}{2}\right)_{\mathrm{AH}} = \rv\frac{\dd \rv}{\dd t} = -\rv\frac{\rv_{0}}{a^{2}} \frac{\dd a}{\dd t} = -H\rv^{2}.
\end{equation}
\noindent
A second term capturing the rate of kinetic energy dissipation is modeled using a parameter $\eta$ that describes the efficiency of the 
energy cascade.
Physically this term would represent
viscosity in the Navier-Stokes equation, but in our simulations it arises from 
dissipative truncation error in the discretization of Euler's equations.
The dissipative term reads
\begin{equation}
\label{eqn:KE_dissipation}
\frac{\dd}{\dd t}\left(\frac{\rv^{2}}{2}\right)_{\mathrm{diss}} = -\eta\frac{\rv^{3}}{aL}.
\end{equation}
\noindent
The relevant length scale in Equation \ref{eqn:KE_dissipation} is the driving scale \citep[e.g.,][]{mac_low1999a},
which is the box size $aL$ in our calculations. 
Simulations of 
driven incompressible 
\citep{gotoh2002a,beresnyak2011a} and transonic \citep{schmidt2006a}
turbulence 
suggest that $\eta\sim1$.
The total rate of change in the turbulent velocity is then
\begin{equation}
\label{eqn:velocity_change}
\frac{\dd \rv}{\dd t} = -H\rv - \eta\frac{\rv^{2}}{aL}.
\end{equation}
\noindent
Noting that $a H = \dd a/\dd t$, and writing the eddy turnover frequency as 
$\omega(\rv,a) \sim \rv / aL$, the 
rate of change of the turbulent velocity with scale factor can be recast as
\begin{equation}
\label{eqn:ah}
\frac{\dd \rv}{\dd a} = -\left( 1 + \eta\frac{\omega}{H}\right) \frac{\rv}{a}.
\end{equation}
\noindent
We will refer to Equation \ref{eqn:ah} as the ``adiabatic heating equation'' 
and it provides quantitative insight into the 
slow and fast contraction regimes discussed qualitatively above.  
It shows that the heating of turbulence during contraction is moderated by 
dissipation with an efficiency proportional to $\sim \omega/|H|$.
When the contraction is slow, the dissipative term 
is larger than the heating term ($\eta\omega/|H|\gg1$) and the velocity
decreases with 
decreasing scale factor ($\dd \rv/ \dd a > 0$). Conversely, when the contraction is fast, 
the velocity increases with decreasing scale factor ($\dd \rv/\dd a < 0$).
Once the eddy turnover and collapse frequencies become comparable and synchronize,
whether the velocities grow or decay depends on how $H$ varies with $a$.

Figure \ref{fig:mach_number} depicts the evolution of the turbulent velocity $\rv$ (upper panels) 
and the ratio of frequencies $\omega/|H|$ (lower panels) obtained from
simulations, along with predictions from the adiabatic heating equation using
a dissipation 
parameter \footnote{We find that using $\eta=1.2$ in Equation \ref{eqn:ah} 
reproduces well the isothermal simulation results at all Mach numbers.}
fixed at $\eta = 1.2$.
The model reproduces well the behavior of both properties of turbulence in contracting gases.
For simulations where $\omega/|H| \gg 1$ or $\omega/|H| \ll 1$ during the 
whole computation (left and 
center panels) the adiabatic heating equation provides an accurate 
description of the turbulent velocity evolution. In the initially fast contraction with constant $H$
(right panel) 
the general behavior is also well modeled by the adiabatic heating equation, but the 
predicted transition from heating to dissipation occurs later in the model than 
in the numerical computation.
These differences may indicate that dissipative effects are delayed in the 
simulation by an eddy turnover time relative to the analytical model.

The model suggests the adiabatic heating mechanism drives the turbulence in the contraction to an 
asymptotic relation between $\omega$ and $|H|$ determined by the scale factor dependence of the
Hubble parameter.  From the adiabatic heating equation and the definition of the
eddy turnover frequency, we obtain
\begin{equation}
\label{eqn:asymptotic_ah}
\frac{d\log(\omega/H)}{d\log(1/a)}=\left(2+\eta\frac{\omega}{H}\right)-\frac{d\log H }{d\log(1/a)}.
\end{equation}
\noindent
The asymptotic relation is approached as $d\log(\omega/H)/d\log(1/a)\to0$.  For
$H=~$constant we expect $\omega/|H|\to2/\eta$ (Simulations A and C), while 
for $H\propto a^{-3/2}$ we have $\omega/|H|\to1/2\eta$ (Simulation B).  
We find that the simulations follow the evolution in $\omega/|H|$ predicted by Equation \ref{eqn:asymptotic_ah} (see Figure 1, lower panels),
but can show substantial run-to-run variations (e.g., Simulation A).

The asymptotic relation between the typical turbulent velocities and scale factor can be deduced from
Equation \ref{eqn:asymptotic_ah}.  By defining
\begin{equation}
\beta \equiv 2 + \frac{d\log H }{d\log a},
\end{equation}
\noindent
from the adiabatic heating equation we find simply 
that $\rv\propto a^{\beta-1}$ once the contraction and eddy turnover frequencies 
have synchronized.

Although the turbulent velocity roughly tracks the expected scaling $\rv\propto a^{-1}$
before an eddy turnover time elapses, 
the degree of adiabaticity depends 
on the small scale dissipation rate. In additional simulations of
contracting incompressible isothermal turbulence 
($M\sim0.05$), we
have found that even low 
resolution simulations display almost exact adiabatic heating. For very large 
isothermal turbulent velocities
($M>10$), we have found that the heating becomes more adiabatic with increasing resolution. 
This sensible behavior does not affect any of the presented 
results which have been tested over a wide range of resolutions (from N=64$^3$ to N=512$^3$) 
and display similar behavior for the same choice of Hubble parameter evolution and initial  
turbulent velocity. 

\section{Discussion}
\label{section:discussion}

Our results have many potential astrophysical applications, but bear
especially on the problem of turbulence in giant molecular clouds (GMCs).
The properties of  GMCs are likely
set by turbulence, as 
the relations between cloud velocity dispersion, size, 
and mass \citep{larson1981a} 
may reflect properties of turbulence
through the velocity structure function 
\citep[see, e.g.,][]{elmegreen2004a}. If so, observations of the 
dispersion-size relation of GMCs 
\citep[e.g.,][]{bolatto2008a,heyer2009a}
roughly agree with the properties of compressible turbulence 
\citep[][]{ballesteros-paredes2006a,kritsuk2007a,federrath2010a}.
It has been argued that 
molecular clouds are in approximate virial equilibrium 
\citep{larson1981a,solomon1987a} such that 
turbulent motions either balance the cloud self-gravity 
or otherwise reflect the depth of the gravitational potential
\citep[e.g.,][]{bertoldi1992a}.
This apparent virial balance poses a significant challenge for 
understanding the origin and evolution of turbulence in molecular clouds, 
as turbulence should dissipate on a 
crossing time \citep[e.g.,][]{goldreich1974a} that is shorter than 
estimates of the cloud lifetime \citep{blitz1980a}. The typical 
GMC lifetime 
is debated because they may not be in exact 
balance or gravitationally bound 
\citep{hartmann2001a,dib2007a,dobbs2011a}, and perhaps undergo frequent 
collisions \citep[e.g.,][]{tasker2009a}.  It remains unclear how 
turbulence could generically provide support against gravitational collapse since without driving it 
quickly dissipates \citep{stone1998a,mac_low1998a,cho2003a}.

Our study suggests that the connection between velocity dispersion and
size may reflect the competition between adiabatic heating and dissipation.
Depending on the nature of the contraction, the adiabatic heating mechanism 
can enable the 
typical turbulent velocity to scale with a positive power of the size of 
a contracting cloud or 
region, preserving a connection between velocity dispersion and cloud
size without an external source for driving the turbulence.  
From the discussion in 
Section \ref{section:model}, the observed scalings of $\rv\propto L^{1/2}$ 
\citep{solomon1987a,heyer2009a}
require $H\propto a^{-1/2}$.  This scaling is quite different than what
might occur in a gravitational collapse, where naively one expects
$H\propto a^{-3/2}$ and $\rv\propto L^{-1/2}$.  If turbulent 
velocities in GMCs do not originate from gravitational collapse, 
adiabatic heating may still provide a method for instilling the
observed scaling relations through other compression mechanisms.

Although our study has focussed on adiabatic heating in 
isotropically contracting turbulence, we have examined other
scenarios.
The inverse process (``adiabatic cooling'') similarly operates in expanding
gases. Using simulations with
a Hubble parameter $H>0$, we have verified that expanding turbulent gases
adiabatically cool if the eddy turnover frequency is less than the
expansion frequency $H$.  If the expansion is 
very rapid ($H\gg\omega$) then the turbulence freezes out with
$\rv\propto a^{-1}$.  Adiabatic cooling may be
relevant for turbulent astrophysical systems that rapidly expand, such as those
formed in high speed impacts or explosions.  We have also simulated
anisotropic systems, and found that adiabatic heating and cooling can operate
simultaneously in different directions depending on the sign of the
effective Hubble parameter for each axis. Applications of anisotropic
compressions include studies of shock--turbulence interaction \citep[e.g.,][]{adams1996a}.

Some previous works presented ideas related to adiabatic heating.
\citet{olson1973a} studied analytically the evolution 
of mean vorticity in incompressible turbulence
in expanding universes without dissipation, and commented that in a contracting
universe the vorticity would ``blow up''.
There were early works exploring whether turbulence could seed structure formation 
\citep[][]{jones1976a,ozernoy1978a} that considered
the adiabatic scaling of velocity with inverse scale 
factor.
The
collisional N-body calculations of \citet{scalo1982a} suggested
that turbulence might slow a gravitational collapse.
\citet{vazquez-semadeni1998a} suggested that turbulent velocites might
depend on the mean density during collapse.
Our description of adiabatic heating has combined and
expanded upon some of these concepts.

\section{Summary}
\label{section:summary}

Using simulations of contracting isothermal turbulent gases
performed with the {\it Athena} code \citep{stone2008a}, we have
identified an ``adiabatic heating'' mechanism by which 
random bulk motions are amplified by compression.  Adiabatic 
heating acts to increase the turbulent velocities of a contracting
gas if the frequency (or Hubble parameter)
$|H|$ of the contraction is larger than
the eddy turnover frequency $\omega \sim \rv /aL$ ($L$ is the initial
box size, $a$ is the scale factor of the contraction, and $\rv$ is the
turbulent velocity).  When $|H|\gg\omega$ the cascade of energy from 
large scales to small scales is limited and dissipation becomes inefficient,
thereby allowing the gas velocities to heat with the adiabatic scaling 
$\rv\propto a^{-1}$ expected from Euler's equations in a contracting background.
When $|H|\ll\omega$, the cascade operates efficiently and energy dissipation
proceeds similarly to turbulent decay in static volumes. In each case, the
turbulent velocities evolve toward $\omega/|H|\sim1$ during contraction.
Using these insights,
we develop an analytical model to describe the rate of change of energy
per unit mass in the gas as a competition between compressive adiabatic heating
and dissipation on small scales.  
The analytical model successfully predicts
the dependence of both the RMS turbulent velocity $\rv$ and the frequency
ratio $\omega/|H|$ on the scale factor $a$.

\acknowledgments

BER is grateful for generous support from the University of Arizona and Steward Observatory.
PMG thanks Jungyeon Cho and Dongsu Ryu for helpful conversations.
The simulations presented in this work were performed on the {\it pangu} cluster at the
California Institute of Technology
Division of Geological and Planetary Sciences.


\end{document}